\documentclass[prl,twocolumn,showpacs,superscriptaddress]{revtex4-1}
\usepackage{amsmath}
\usepackage{graphicx}
\usepackage{times}

\newcount\minute
\newcount\hour
\newcount\hourMins
\def\now
{
  \minute=\time
  \hour=\time \divide \hour by 60
  \hourMins=\hour \multiply\hourMins by 60
  \advance\minute by -\hourMins
  \zeroPadTwo{\the\hour}:\zeroPadTwo{\the\minute}
}\def\timestamp
{
  \today\ \now
}
\def\today
{
  \zeroPadTwo{\the\day}-\zeroPadTwo{\the\month}-\the\year%
}
\def\zeroPadTwo#1%
{
  \ifnum #1<10 0\fi
  #1%
}
\def \dif {{d}}

\date{\timestamp}

\begin{document}

\title{Excitation spectrum of the Lieb-Liniger model}

\author{Zoran Ristivojevic}
\affiliation{Centre de Physique Th\'{e}orique, Ecole Polytechnique, CNRS, 91128 Palaiseau, France}

\begin{abstract}
We study the integrable model of one-dimensional bosons with contact repulsion. In the limit of weak interaction, we use the microscopic hydrodynamic theory to obtain the excitation spectrum. The statistics of quasiparticles changes with the increase of momentum. At lowest momenta good quasiparticles are fermions, while at higher momenta they are Bogoliubov bosons, in accordance with recent studies. In the limit of strong interaction, we analyze the exact solution and find exact results for the spectrum in terms of the asymptotic series. Those results undoubtedly suggest that fermionic quasiparticle excitations actually exist at all momenta for moderate and strong interaction, and also at lowest momenta for arbitrary interaction. Moreover, at strong interaction we find highly accurate analytical results for several relevant quantities of the Lieb-Liniger model.
\end{abstract}
\pacs{67.85.De, 71.10.Pm}

\maketitle

Recent experimental achievements with ultracold atoms \cite{bloch+08RevModPhys} urge us to deeply understand properties of interacting bosonic systems. One of the most fundamental problems is the excitation spectrum. For weakly interacting bosons in three dimensions, it is given by the Bogoliubov theory \cite{bogoliubov47}. The spectrum is generically linear at lowest momenta, but becomes quadratic above a certain interaction-dependent crossover scale, consistent with observations in cold atomic gases \cite{katz+02PhysRevLett.88.120407}.

The behavior of interacting bosons in lower dimensions is particular due to the pronounced role of quantum fluctuations \cite{Mermin+66,hohenberg67}, preventing us from directly applying the Bogoliubov theory. Hopefully, the existence of integrable models in one dimension enables us to understand more. An example is the Lieb-Liniger model, which describes bosons with contact repulsion \cite{lieb1963exact}. It is remarkable not only due  to the existence of an exact solution but also to the possibility of realization with cold atomic gases \cite{kinoshita+04tonksgas,paredes+04tonksgas}. Furthermore, the model has two branches of elementary excitations. One is the conventional type I excitation branch that exhibits the same features predicted by Bogoliubov theory in the limiting cases of small and large momenta \cite{girardeau60,lieb1963excitations}. Second, the type II branch describes solitonlike excitations \cite{lieb1963excitations,stringaribook}.

The universal low-energy theory of interacting quantum particles in one dimension is usually described by the paradigm of the Luttinger liquid. The excitations in this theory are phonons. These bosonic quasiparticles represent the waves of particle density propagating with constant velocity \cite{Giamarchi}. Despite its tremendous success \cite{Giamarchi}, Luttinger liquid description is only the limiting theory, since its foundation is the linear spectrum. This is clearly an approximation for more realistic models. Recent efforts to understand effects beyond the linear theory have yielded the picture of fermionic quasiparticles that are true low-energy excitations \cite{rozhkov2005fermionic,imambekov+12RevModPhys.84.1253}  in one-dimensional quantum liquids. This is an interesting result, different from the conventional picture of phonon excitations in a Luttinger liquid.

In this Letter we consider the Lieb-Liniger model. Using the microscopic theory and the exact solution, we find the excitation spectrum at arbitrary momenta and in a wide region of interaction strengths. Our microscopic results fully support the phenomenological picture of fermionic quasiparticles that are lowest-energy excitations at arbitrary interaction strength \cite{imambekov+12RevModPhys.84.1253}. Bosonic quasiparticles that are characterized by the Bogoliubov form of the spectrum exist only at weak interaction and at higher momenta.

We study the system of interacting bosons described by the Lieb-Liniger model \cite{lieb1963exact},
\begin{align}\label{Horiginal}
H=-\frac{\hbar^2}{2m}\sum_{i=1}^N \frac{\partial^2}{\partial x_i^2}+\frac{\hbar^2 c}{2m}\sum_{i\neq j} \delta(x_i-x_j).
\end{align}
Here $m$ is the mass of particles, while $c>0$ describes the repulsion strength. We consider the thermodynamic limit $N,\,L\to\infty$, such that the mean density $n=N/L$ is finite. For convenience, we introduce the dimensionless parameter $\gamma=c/n$ that characterizes the interaction strength. In the following we primarily consider the type I branch of elementary excitations.

In the regime of weak interaction, $\gamma\ll 1$, we use the hydrodynamic approach \cite{popov72,haldane81prl} to study the model (\ref{Horiginal}). We start from the standard expression for the Hamiltonian of interacting bosons in second quantization \cite{stringaribook} and reexpress the bosonic single particle operator as $\psi^\dagger(x)=\sqrt{n(x)}e^{i\theta(x)}$, where $n(x)$ and  $\theta(x)$ are the fluctuating bosonic density and the phase fields, respectively. Accounting for small density fluctuations in the standard way \cite{cazalilla+2011RMP,haldane81prl}, where $n(x)=n+\nabla\varphi(x)/\pi$, we obtain
\begin{align}\label{H}
H=&\frac{\hbar^2}{2m}\int\dif x\left[\left(n+\frac{\nabla\varphi}{\pi}\right) (\nabla\theta)^2+\frac{(\nabla^2\varphi)^2}{4\pi^2n}\right]\notag\\ &+\frac{\hbar^2 c}{2\pi^2 m}\int\dif x (\nabla\varphi)^2.
\end{align}
The fields $\theta$ and $\varphi$ satisfy the commutation relation $[\nabla\varphi(x),\theta(y)]=-i\pi\delta(x-y)$. The Hamiltonian (\ref{H}) provides an effective description of the original one, given by Eq.~(\ref{Horiginal}), at momenta below $\hbar n$. In this regime, the fluctuations of the field $\nabla\varphi$ are small, enabling us to use the hydrodynamic approach.

At lowest momenta, the excitation spectrum is determined by the most relevant operators of the Hamiltonian (\ref{H}). Retaining the operators of scaling dimension two, $(\nabla\varphi)^2$ and $(\nabla\theta)^2$, we obtain the Luttinger liquid Hamiltonian,  $H_0$. It describes the excitations with linear spectrum $\varepsilon_p=v p$, where $v=(\hbar n/m)\gamma^{1/2}$ is the sound velocity \cite{lieb1963exact}. By $p$ we denote the momentum. It is important to note that the Luttinger liquid Hamiltonian does not uniquely determine statistics of quasiparticle excitations. Indeed, $H_0$ exactly describes both, the excitations in a system of noninteracting (i) bosons \cite{haldane81prl} and (ii)  fermions \cite{mattis+65} with linear dispersion. However, the theory (\ref{H}) has operators of higher scaling dimension that arise from amplitude fluctuations of $\psi$. They lift the statistics degeneracy of $H_0$, and thus uniquely determine true quasiparticles. At lowest momenta we must include the leading irrelevant operator, which is the one of scaling dimension three. The resulting Hamiltonian  $H_0+(\hbar^2/2\pi m)\int\dif x (\nabla\varphi)(\nabla\theta)^2$ is diagonalized by the fermionization procedure. We obtain the low-energy spectrum
\begin{align}\label{E_weakint_fermions}
\varepsilon_p=vp+\frac{p^2}{2m^*}
\end{align}
with the quasiparticle mass $m^*=4\pi^{1/2}m/3\gamma^{1/4}$. The quadratic dispersion (\ref{E_weakint_fermions}) is in agreement with the earlier result obtained phenomenologically \cite{imambekov+12RevModPhys.84.1253}, and very recent result \cite{pustilnik+14} obtained by studying the exact solution of the Lieb-Liniger model at weak interaction. However, unlike in the latter study, our microscopic theory directly identifies fermionic nature of quasiparticle excitations at lowest momenta.

At higher momenta, operators of higher scaling dimension could be more important than the ones of lower dimension. Omitting the operator of scaling dimension three from the Hamiltonian (\ref{H}) enables us to diagonalize it. We obtain that it describes bosonic quasiparticles with  Bogoliubov spectrum
\begin{align}\label{E_bogoliubov}
\varepsilon_p=v p\sqrt{1+\frac{p^2}{4m^2v^2}}.
\end{align}
Comparing the first subleading term of the two spectra (\ref{E_weakint_fermions}) and (\ref{E_bogoliubov}), we infer the crossover momentum scale \cite{imambekov+12RevModPhys.84.1253} $p^*=\hbar n\gamma^{3/4}$. At $p\ll p^*$, the operator of scaling dimension three cannot be neglected. In this regime we find the fermionic quasiparticles with the spectrum (\ref{E_weakint_fermions}). At $p\gg p^*$, the operator of scaling dimension four in Eq.~(\ref{H}) is the leading correction to $H_0$, yielding the spectrum (\ref{E_bogoliubov}) of bosonic quasiparticles. In this regime the neglected operator of scaling dimension three (and many others of higher scaling dimension) describe residual interaction of Bogoliubov quasiparticles, which is  responsible, e.g., for the shape of the spectral function that has nondelta function shape \cite{khodas+07bosonsPhysRevLett.99.110405}.

At momenta below the crossover momentum $p_0=mv$, the bosonic dispersion (\ref{E_bogoliubov}) simplifies into
\begin{align}
\varepsilon_p=vp+\frac{p^3}{8m^2 v},
\end{align}
and describes Bogoliubov phonons. Reexpressing the asymptote as $p_0=\hbar n\gamma^{1/2}$, we observe that $p_0$ and $p^*$, when extrapolated to moderate interaction, cross each other at $\gamma_c=1$. If such extrapolation from the weakly interacting region $\gamma\ll 1$ indeed holds, this would imply limited parameter regime where phonon quasiparticles exist. Moreover, the bosonic quasiparticles from the region above $p_0$, which have the spectrum \cite{Note1,kulish+1976comparison} 
\begin{align}\label{Ehighp}
\varepsilon_p=\frac{p^2}{2m}+\gamma\frac{\hbar^2n^2}{m},
\end{align}
are also expected to cease together with the phonons as interaction strength is increased. In order to address those questions, in the following we calculate the excitation spectrum of the model (\ref{Horiginal}) at strong interaction, $\gamma\gg 1$. We note that in this regime the microscopic approach (\ref{H}) is inapplicable.

The model (\ref{Horiginal}) is solvable by Bethe ansatz \cite{lieb1963exact,lieb1963excitations,korepin1993book}. This technique produces a set of equations that contains all the information about the excitation spectrum. However, extracting it explicitly is in general an involved task. For type I excitations, the spectrum is implicitly given by
\begin{align}\label{pE}
p=2\pi\hbar Q\int_{1}^{k/Q}\dif x \rho(x),\quad \varepsilon=\frac{\hbar^2 Q^2}{m}\int_{1}^{k/Q}\dif x \sigma(x),
\end{align}
in terms of the parameter $k$. It satisfies $k>Q$, where $Q$ is the Fermi rapidity, defined by the condition
\begin{align}\label{Q}
Q\int_{-1}^{1}\dif x \rho(x)=n.
\end{align}
The two density functions in Eq.~(\ref{pE}) can be expressed as
\begin{align}\label{rhosigma}
\rho(x)=\frac{f(x)+f(-x)}{2},\quad \sigma(x)=\frac{f(x)-f(-x)}{2},
\end{align}
in terms of the solution of the integral equation
\begin{align}\label{f}
f(x)-\frac{1}{\pi}\int_{-1}^{1} \dif y \frac{\lambda}{\lambda^2+(x-y)^2}f(y)=\frac{1}{2\pi}+x.
\end{align}
The function $f(x)$, and thus $\rho(x)$ and $\sigma(x)$, are defined for any real $x$, and depend on the dimensionless parameter $\lambda=c/Q$.

In the Tonks-Girardeau limit, $\gamma\to \infty$, and thus  $\lambda\to \infty$, so Eq.~(\ref{f}) becomes trivial. The solution is $\rho=1/2\pi$ and $\sigma=x$. From Eqs.~(\ref{pE}) and (\ref{Q}) we then find $p=\hbar(k-Q)$, $\varepsilon=(\hbar^2/2m)(k^2-Q^2)$, and $Q=\pi n$, which yields \cite{girardeau60,lieb1963excitations}
\begin{align}\label{Etonks}
\varepsilon_p=\frac{\pi\hbar n}{m}p+\frac{p^2}{2m}.
\end{align}
Equation (\ref{Etonks}) is exact for any $p$ and describes quasiparticle excitations in the system of bosons with contact interaction of infinite strength. In this case the repulsion prevents two bosons to share the same space position, acting effectively as Pauli principle on fermions.

At moderate and strong interaction, we solve Eq.~(\ref{f}) by the power series method. We assume the solution in the form
\begin{align}\label{fassumption}
f(x)=\sum_{n=0}^{\infty} a_n P_n(x),
\end{align}
where $P_n$ are Legendre polynomials and the coefficients $a_n$ are to be determined. At $\lambda>|x-y|$, we expand the kernel of Eq.~(\ref{f}) into power series. It is convenient to introduce $F_n^\ell=\int_{-1}^{1} \dif x\, x^\ell P_n(x)$. This expression is nonzero only at $\ell\ge n$ provided $\ell+n$ is an even integer, and then it becomes \cite{gradshteyn}
\begin{align}
F_n^\ell=\frac{2^{n+1}\ell !\left(\frac{\ell+n}{2}\right)!} {(\ell+n+1)!\left(\frac{\ell-n}{2}\right)!}.
\end{align}
Using the orthogonality of Legendre polynomials, we obtain the relations between the coefficients of $f(x)$,
\begin{align}\label{set}
\frac{2a_n}{2n+1}=&\sum_{m=0}^{M} \sum_{\ell=0}^{2m}\sum_{r=0}^{\ell} \frac{(-1)^{m+\ell}(2m)!}{\pi\ell!(2m-\ell)!} \frac{a_r}{\lambda^{2m+1}} F_r^\ell F_n^{2m-\ell}\notag\\
&+\frac{1}{\pi}\delta_{n,0} +\frac{2}{3}\delta_{n,1},\quad M\to \infty.
\end{align}
This is an infinite set of linear equations that determines $a_n$ in Eq.~(\ref{fassumption}). Since $r\le 2m$ in the summation in Eq.~(\ref{set}), all the coefficients $a_n$ for $n\ge 2$ scale at least as fast as  $\lambda^{-n-1}$ at large $\lambda$ \cite{Note2}. 
This enables us to systematically solve Eq.~(\ref{set}) at finite $M$, which makes finite set of equations for $\{a_0,a_1,\ldots, a_{2M}\}$. Likewise $\rho(x)$ and $\sigma(x)$, Legendre polynomials have parity. Therefore, only the polynomials $P_n$ with $n$ being even (odd) participate in the series for $\rho(x)$ ($\sigma(x)$).

At $M=0$, Eq.~(\ref{set}) leads to the spectrum as in Eq.~(\ref{Etonks}) multiplied by an overall factor of $1-4/\gamma+12/\gamma^2$. Already $M=1$ is sufficient to obtain nontrivial corrections to the spectrum (\ref{Etonks}), which start at order $\mathcal{O}(1/\gamma^3)$. From the solution of Eq.~(\ref{f}), we obtain
\begin{align}\label{a0}
\rho(x)=&\frac{1}{2\pi}+ \frac{1}{\pi^2\lambda} + \frac{2}{\pi^3\lambda^2} + \frac{12-\pi^2}{3\pi^4\lambda^3} +\frac{8-2\pi^2}{\pi^5\lambda^4}\notag\\ &-{x^2}\left(\frac{1}{\pi^2\lambda^3} +\frac{2}{\pi^3\lambda^4}\right) +\mathcal{O}(\lambda^{-5}),\\
\label{a1a2} \sigma(x)=&x\left(1+\frac{4}{3\pi\lambda^3} \right)+\mathcal{O}(\lambda^{-5}).
\end{align}
Using Eq.~(\ref{Q}) we now find the inverse series $\lambda=(\gamma+2)/\pi -4\pi/3\gamma^2+16\pi/3\gamma^3+\mathcal{O}(\gamma^{-4})$, while $Q=\gamma n/\lambda$ can be easily expressed as a function of $\gamma$. Finally, elementary algebra with Eq.~(\ref{pE}) leads to the low-momentum dispersion
\begin{align}\label{E-fermions-low}
\varepsilon_p=&vp+\frac{p^2}{2m^*} +\left(\frac{8\pi}{3\gamma^3}-\frac{80\pi}{3\gamma^4} \right)\frac{p^3}{\hbar n m}\notag\\ &+\left(\frac{2}{3\gamma^3}-\frac{20}{3\gamma^4}\right) \frac{p^4}{\hbar^2 n^2 m} +\mathcal{O}\left(\gamma^{-5}\right).
\end{align}
Here the sound velocity is given by $v=\pi\hbar n/mK$, where the Luttinger liquid parameter takes the value
\begin{align}
K=1+\frac{4}{\gamma}+\frac{4}{\gamma^2} -\frac{16\pi^2}{3\gamma^3}+ \frac{32\pi^2}{3\gamma^4}+\mathcal{O}(\gamma^{-5}).
\end{align}
The mass is given by $m/m^*=1-4/\gamma+12/\gamma^2+(32\pi^2-96)/3\gamma^3 -(320\pi^2-240)/3\gamma^4 +\mathcal{O}(\gamma^{-5})$. This result is in agreement with the exact relation for the model (\ref{Horiginal}),
\begin{align}\label{m*}
\frac{m}{m^*} =(1-\gamma\partial_\gamma) \frac{1}{\sqrt{K}},
\end{align}
which could be obtained from a more general expression one finds in Refs.~\cite{pereira+2006PRL,imambekov+12RevModPhys.84.1253}.

The spectrum (\ref{E-fermions-low}) contains the leading order terms in the asymptotic expansion at momenta $p\ll p_0^*$, where $p_0^*=\hbar n\gamma$. The origin for this limitation is the condition on $\lambda$ that we required in order to expand the kernel of the integral equation (\ref{f}). We note that the quantities that involve $\rho(x)$ and $\sigma(x)$ only for $|x|<1$, such as the ground state energy or the spectrum of type II excitations, in principle do not have any limitation already at $\lambda>2$ (corresponding to $\gamma>4.527$), provided one solves Eq.~(\ref{set}) at arbitrary $\lambda$. However, here we solved Eq.~(\ref{set}) using a large $\lambda$ expansion. We have verified that the set of equations (\ref{set}) when solved for $M>1$, only leads to changes in Eqs.~(\ref{a0}) and (\ref{a1a2}) and all subsequent expression starting from the order $\mathcal{O}(\gamma^{-5})$. In Supplemental Material \cite{Note2} we give analytical expressions of relevant quantities for the Lieb-Liniger model at high order.

At sufficiently high momenta, the upper limit of integration in Eq.~(\ref{pE}) exceeds $\lambda$, and therefore for so high momenta the expansion (\ref{fassumption}) cannot be used. Instead, we find the solution of Eq.~(\ref{f}) by iterations, which leads to
\begin{gather}\label{rhoiteration}
\rho(x)=\frac{1}{2\pi}\left[1+\frac{\lambda} {\arctan\lambda}\frac{1}{ (x^2+\lambda^2) }\right],\\
\label{sigmaiteration}
\sigma(x)=x\left(1 +\frac{\arctan\frac{2\lambda}{\lambda^2+x^2-1}} {\pi}\right)+\frac{\lambda}{2\pi}\ln\frac{(1-x)^2+\lambda^2} {(1+x)^2+\lambda^2}.
\end{gather}
We easily verify that Eq.~(\ref{sigmaiteration}) at large $\lambda$ is in agreement with the result (\ref{a1a2}). On the other hand, Eq.~(\ref{rhoiteration}) agrees with Eq.~(\ref{a0}) to order $\mathcal{O}(\lambda^{-2})$. We have checked that the next iteration of Eq.~(\ref{rhoiteration}) in the initial integral equation produces such $\rho(x)$ that agrees with Eq.~(\ref{a0}) to order $\mathcal{O}(\lambda^{-3})$. Therefore, the low momentum spectrum (\ref{E-fermions-low}) could be also reproduced from the densities obtained by the iteration procedure.

Using the density (\ref{rhoiteration}) in Eq.~(\ref{pE}), we obtain
\begin{align}\label{plarge}
p=\hbar k-\hbar Q\frac{\arctan\frac{\lambda Q}{k}}{\arctan\lambda}.
\end{align}
At high $p$, the parameter $k$ is also large, and thus the second term of Eq.~(\ref{plarge}) can be neglected. Furthermore, $\varepsilon$ from Eq.~(\ref{pE}) in the leading order is determined from the region of large $x$, where Eq.~(\ref{sigmaiteration}) becomes $\sigma(x)=x+\mathcal{O}(x^{-3})$. Therefore, in the leading order we obtain the expected result at high momenta, $\varepsilon_p=p^2/2m$. A more careful way is to substitute Eq.~(\ref{sigmaiteration}) in Eq.~(\ref{pE}), resulting in
\begin{align}\label{Elarge}
\varepsilon=&\frac{\hbar^2(k^2-Q^2)}{2m} \left(1+\frac{\arctan\frac{2\lambda} {\chi^2+\lambda^2-1}}{\pi}\right) \notag\\
&+ \frac{\hbar^2n^2\gamma^2}{2\pi m}\biggl[\arctan\frac{2\chi^2-2}{\lambda(\chi^2 +\lambda^2+3)}+ \frac{\ln\left(1+\frac{4}{\lambda^2}\right)}{\lambda}\notag\\ &+\frac{\chi}{\lambda}\ln\frac{(\chi-1)^2+\lambda^2} {(\chi+1)^2+\lambda^2}\biggr],\qquad \chi=\frac{k}{Q}.
\end{align}
In the limiting case of high momenta, $p\gg p_0^*$, Eqs.~(\ref{plarge}) and (\ref{Elarge}) give
\begin{align}\label{Elargegamma}
\varepsilon_p=\frac{p^2}{2m}+2\gamma \frac{\hbar^2n^2}{m}-\frac{\pi^2\hbar^2 n^2}{2m}+\mathcal{O}\left(\frac{1}{p^2}\right),
\end{align}
in agreement with the original work \cite{lieb1963excitations}. However, the analytical expressions (\ref{plarge}) and (\ref{Elarge}) are the dispersion relation of the Lieb-Liniger model, at arbitrary momentum and for $\lambda$ being moderate and large. In Fig.~1 we show the full dispersion curve and compare it with the limiting expressions (\ref{E-fermions-low}) and (\ref{Elargegamma}).

\begin{figure}
\includegraphics[width=0.8\columnwidth]{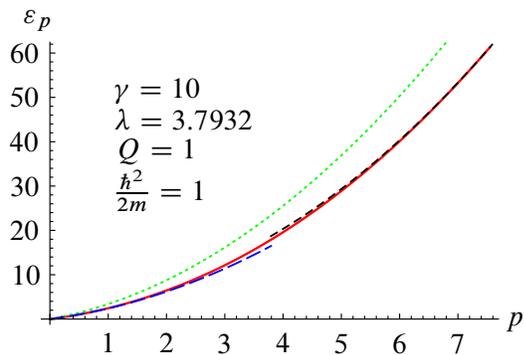}
\caption{The dispersion curve for the Lieb-Liniger model at $\gamma=10$ is represented by solid line, see Eqs.~(\ref{plarge}) and (\ref{Elarge}). Dashed lines are the two limiting cases, Eqs.~(\ref{E-fermions-low}) and (\ref{Elargegamma}). Doted line is given for comparison, and represents the limiting spectrum at $\gamma\to\infty$, Eq.~(\ref{Etonks}). In the units used for the plot, $p_0^*=\lambda$.}\label{fig1}
\end{figure}

\begin{figure}
\includegraphics[width=0.8\columnwidth]{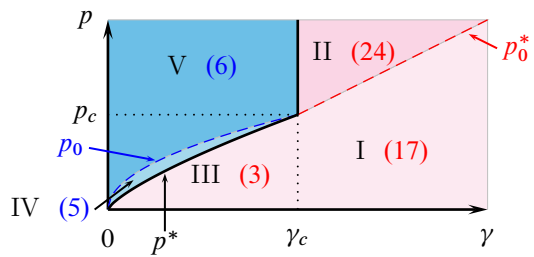}
\caption{The spectrum of elementary excitations in the Lieb-Liniger model has its characteristic form in five regions of the momentum-interaction plane. The number in parentheses denotes the number of equation where the limiting form of the spectrum is given. In regions I, II, and III, elementary excitations are fermionic quasiparticles, while IV and V they are bosonic quasiparticles. The crossover momenta are $p^*\sim\gamma^{3/4}$, $p_0\sim\gamma^{1/2}$, and $p_0^*\sim\gamma$, see the main text. As we neglected the pure numbers in all the asymptotes, $\gamma_c=1$, corresponding to $p_c=\hbar n$.}\label{fig2}
\end{figure}

We are now in position to distinguish the excitation spectrum of the Lieb-Liniger model in several regions of the momentum-interaction plane, see Fig.~\ref{fig2}. At smallest momenta, corresponding to regions I and III, the system is characterized by a linear dispersion with the quadratic correction controlled by the mass $m^*$. This is in agreement with earlier phenomenological and Bethe ansatz studies \cite{imambekov+12RevModPhys.84.1253,pustilnik+14}. As follows from our microscopic theory (\ref{H}), at weak interaction the low-momentum excitations are fermionic quasiparticles. It is quite expected that the same quasiparticles extend to region I, since the two spectra (\ref{E_weakint_fermions}) and (\ref{E-fermions-low}) have the same form \cite{Note3}. 
The mass $m^*$ depends on the interaction strength as given by Eq.~(\ref{m*}). At momenta above $p_0^*$, the quasiparticle mass decreases, becoming the bare mass of original particles, $m$. We note that this change of the mass is accompanied with the disappearance of the linear term in the dispersion, cf. Eqs.~(\ref{E-fermions-low}) and (\ref{Elargegamma}).

At higher momenta, the cubic correction in the spectrum may become dominant with respect to the quadratic one. This indeed occurs at weak interaction, where fermionic quasiparticles (region III) transform into Bogoliubov phonons (region IV) as the momentum is increased \cite{pustilnik+14}. At strong interaction, such change in the nature of quasiparticles does not occur. As follows from Eq.~(\ref{E-fermions-low}), the crossover momentum where this would happen is at momenta above $p_0^*$, deeply in region II. Thus, at strong interaction there is no space for bosonic quasiparticles to develop. The only possibility is to have the dispersion crossover region between quasiparticles of the same statistics, which happens at $p_0^*$.

We finally note that Bogoliubov quasiparticles at high momenta exist at weak interaction, see region V in Fig.~\ref{fig2}. Those quasiparticles are expected to exist as long as Bogoliubov phonons at lower momenta are present. The first subleading term of the quasiparticle dispersion in regions V and II are different, cf. Eqs.~(\ref{Elargegamma}) and (\ref{Ehighp}). This is consistent with the picture of fermionic quasiparticles at strong interaction, which due to Pauli principle are expected to have higher energy than the bosonic ones of the same momentum.

The excitation spectrum of a three-dimensional Bose-Einstein condensate is measured by Bragg spectroscopy \cite{stenger+99} in both limits of weak \cite{katz+02PhysRevLett.88.120407} and strong interaction \cite{papp+08}. In principle, the same technique could be used to probe the spectrum of the Lieb-Liniger model. Knowledge of the spectrum can be used to find the singularities in the dynamic structure factor \cite{imambekov+12RevModPhys.84.1253}. It is peaked and diverges in the vicinity of type I excitation branch as a power law, with different exponents in the fermionic and bosonic regions (see Fig.~\ref{fig2}). We finally notice that in the presence of integrability breaking perturbations (always present in realistic conditions), this peak becomes broadened. Such effect is intimately connected to the quasiparticle decay rate, which has different scaling with momentum in the fermionic \cite{matveev+13PhysRevLett.111.256401} and the phononic regions \cite{ristivojevic+14PhysRevB.89.180507}. This is yet another manifestation of the change of quasiparticle statistics which could be tested by Bragg spectroscopy \cite{fabbri+11PhysRevA.83.031604}.

The power series method could be used to obtain other relevant quantities of the Lieb-Liniger model. For example, the ground state energy can be expressed as $E_0=(\hbar^2 n^2 N/2m)e(\gamma)$, where  $e(\gamma)=(\gamma^3/\lambda^3) \int_{-1}^{1}\dif x x^2\rho(x)$, see Ref.~\cite{lieb1963exact}. Using Eq.~(\ref{a0}), we find
$
e(\gamma)={\pi^2}/{3}-{4\pi^2}/{3\gamma} +{4\pi^2}/{\gamma^2} -{32\pi^2 (15-\pi^2)}/{45\gamma^3}
-{16\pi^2(4\pi^2-15)}/{9\gamma^4} +\mathcal{O}(\gamma^{-5})
$. The first four terms in $e(\gamma)$ are in agreement with Ref.~\cite{guan2011polylogs}. At strong interaction, from the spectrum (\ref{E-fermions-low}) of type I excitations one directly obtains the spectrum type II excitations as $\tilde{\varepsilon}_p=-\varepsilon_{-p}$. The latter excitations branch is recently studied in Ref.~\cite{Astrakharchik+13}.

In conclusion, we calculated the excitation spectrum of the Lieb-Liniger model using the microscopic theory and the exact solution. At arbitrary interaction, the quasiparticle excitations at small momenta are characterized by the mass $m^*$ that is always larger than the mass of original particles. Our results give strong evidence that fermionic quasiparticles actually exist in a huge parameter range. Bosonic Bogoliubov quasiparticles are present only at weak interaction.
Understanding the physics near the boundary region between the fermionic and bosonic quasiparticles, see Fig.~\ref{fig2}, is a challenging problem. Our results have direct application to the dual Cheon-Shigehara model \cite{cheon+PhysRevLett.82.2536,yukalov,khodas+07bosonsPhysRevLett.99.110405}.

I thank K.~A.~Matveev for helpful discussions and L.~I.~Glazman for a correspondence. This work was supported by PALM Labex.


%

\onecolumngrid
\newpage
\setcounter{equation}{0}
\setcounter{figure}{0}

\renewcommand{\theequation}{S\arabic{equation}}
\renewcommand{\thepage}{S\arabic{page}}
\renewcommand{\thesection}{S\arabic{section}}
\renewcommand{\thetable}{S\arabic{table}}
\renewcommand{\thefigure}{S\arabic{figure}}
\renewcommand{\bibnumfmt}[1]{[{\normalfont S#1}]}
\renewcommand{\citenumfont}[1]{S#1}

\begin{center}
{\large\textbf{Excitation spectrum of the Lieb-Liniger model}
\\\vskip 5pt
Supplemental material
}
 \end{center}

\begin{center}
Zoran Ristivojevic$^{1}$
\\
\vskip 1mm
\textit{$^1$Centre de Physique Th\'{e}orique, Ecole Polytechnique, CNRS, 91128 Palaiseau, France}
\end{center}
\vskip 5mm

Here we present some details about the structure of solutions of Eq.~(14). It enables us to find highly accurate results for several quantities of the Lieb-Liniger model at strong interaction.

At large $\lambda$, from Eq.~(14) we infer that the coefficients $a_n(\lambda)$ defined in Eq.~(12),  scale as $a_n\sim \lambda^{-n-1}$ for even $n\ge 2$, while the ones having an odd index $n\ge 3$, scale as  $a_n\sim \lambda^{-n-2}$. For $M=0$, we can correctly obtain only the first three terms in the expansion of Eqs.~(15) and (16). We easily find
\begin{align}\label{aaa}
a_0=\frac{1}{2\pi}+\frac{1}{\pi^2\lambda} +\frac{2}{\pi^3\lambda^2}+\mathcal{O}(\lambda^{-3}),\quad a_1=1.
\end{align}
For $M=1$, we obtain an additional coefficient
\begin{align}\label{aa}
a_2=-\frac{2}{3\pi^2\lambda^3}-\frac{4}{3\pi^3\lambda^4} +\mathcal{O}(\lambda^{-5}).
\end{align}
However, at $M=1$ we must use such $a_0$ and $a_1$ that include higher order terms in their expansion,
\begin{align}\label{aaaa}
a_0=\frac{1}{2\pi}+\frac{1}{\pi^2\lambda} +\frac{2}{\pi^3\lambda^2} -\frac{2(\pi^2-6)}{3\pi^4\lambda^3} -\frac{8(\pi^2-3)}{3\pi^5\lambda^4} +\mathcal{O}(\lambda^{-5}),\quad a_1=1+\frac{4}{3\pi\lambda^3}+\mathcal{O}(\lambda^{-5}).
\end{align}
It is very important to note that the newly added terms in $a_0$ and $a_1$, which are proportional to $\lambda^{-3}$ and $\lambda^{-4}$, do not affect the ones obtained at smaller $M$, cf. Eqs.~(\ref{aaa}) and (\ref{aaaa}). For $M=2$ one obtains two more coefficients, $a_3\sim \mathcal{O}(\lambda^{-5})$ and $a_4\sim\mathcal{O}(\lambda^{-5})$, and thus one must use $a_0$, $a_1$, and $a_2$ with higher accuracy, including the orders $\mathcal{O}(\lambda^{-5})$ and $\mathcal{O}(\lambda^{-6})$. For $M=3$,  one obtains two more coefficients, $a_5\sim \mathcal{O}(\lambda^{-7})$ and $a_6\sim \mathcal{O}(\lambda^{-7})$, and uses the previous ones with higher accuracy, and so on.

Let us consider, for example, the Luttinger liquid parameter $K$. At $M=0$ we obtain $\lambda=(\gamma+2)/\pi$, which leads to the first three terms in the expansion, $K=1+4/\gamma+4/\gamma^2$. At $M=1$ we obtain two more terms, $-16\pi^2/3\gamma^3+32\pi^2/3\gamma^4$, as given in Eq.~(18) of the main text. At $M=2$ we obtain two more terms that are higher order in $1/\gamma$ than the ones of Eq.~(18). We notice that whenever we solve Eq.~(14) at $M+1$, having, e.g., $K$ calculated at $M$, we add a new Legendre polynomial in the expression for $\rho(x)$ that brings two more powers of $\pi$ in the numerators of newly added terms.

In the following we give highly accurate expression for several quantities of the Lieb-Liniger model. They are obtained by solving Eq.~(14) at $M=3$ in the limit of large $\lambda$. Using the notation from the main text, we find the expressions displayed below, which  include first nine terms of the asymptotic expansion. One could easily calculate more terms in the expansion by using $M>3$. However, higher order terms become progressively longer, which is the reason of not displaying them here. We obtain
\begin{align}\label{Kprecise}
K=&1+ \frac{4}{\gamma } +\frac{4}{\gamma ^2} -\frac{16 \pi ^2}{3 \gamma ^3} +\frac{32 \pi ^2}{3 \gamma ^4} +\frac{64 \pi ^2 \left(-5+3 \pi
   ^2\right)}{15 \gamma ^5} -\frac{32 \pi ^2 \left(-20+39 \pi
   ^2\right)}{15 \gamma ^6}\notag\\
    &-\frac{256 \pi ^2 \left(35-147
   \pi ^2+15 \pi
   ^4\right)}{105 \gamma ^7}
 +\frac{128 \pi ^2
   \left(140-1050 \pi ^2+341
   \pi ^4\right)}{105 \gamma
   ^8} +\mathcal{O}(\gamma^{-9}),
\end{align}

\begin{align}\label{mprecise}
\frac{m}{m^*}=&1 -\frac{4}{\gamma } +\frac{12}{\gamma ^2} +\frac{32 \left(-3+\pi^2\right)}{3 \gamma ^3} -\frac{80 \left(-3+4 \pi
   ^2\right)}{3 \gamma ^4} +\frac{64 \left(-15+50 \pi ^2-3
   \pi ^4\right)}{5 \gamma ^5} +\frac{112 \left(12-80 \pi
   ^2+17 \pi ^4\right)}{3
   \gamma ^6}\notag\\& -\frac{256 \left(420-4900 \pi
   ^2+2401 \pi ^4-60 \pi
   ^6\right)}{105 \gamma ^7} +\frac{192 \left(420-7840 \pi
   ^2+7252 \pi ^4-605 \pi
   ^6\right)}{35 \gamma ^8} +\mathcal{O}(\gamma^{-9}),
\end{align}

\begin{align}
\frac{Q}{\pi n}=&1 -\frac{2}{\gamma } +\frac{4}{\gamma ^2} +\frac{4 \left(-6+\pi
   ^2\right)}{3 \gamma ^3} -\frac{16 \left(-3+2 \pi
   ^2\right)}{3 \gamma ^4} +\frac{32 \left(-15+25 \pi
   ^2-\pi ^4\right)}{15 \gamma
   ^5} +\frac{32 \left(90-300 \pi
   ^2+43 \pi ^4\right)}{45
   \gamma ^6}\notag\\& -\frac{32 \left(252-1470 \pi
   ^2+490 \pi ^4-9 \pi
   ^6\right)}{63 \gamma ^7} +\frac{128 \left(90-840 \pi
   ^2+532 \pi ^4-33 \pi
   ^6\right)}{45 \gamma ^8} +\mathcal{O}(\gamma^{-9}),
\end{align}

\begin{align}
\lambda=&\frac{\gamma}{\pi}+\frac{2}{\pi} -\frac{4 \pi }{3 \gamma ^2} +\frac{16 \pi }{3 \gamma ^3} +\frac{16\pi \left(-15+2 \pi ^2
   \right)}{15 \gamma ^4} -\frac{16\pi \left(-40+19 \pi ^2\right)}{15 \gamma ^5}  +\frac{32\pi \left(-70 +77 \pi
   ^2-3 \pi ^4\right)}{21
   \gamma ^6}\notag\\
   &+\frac{64\pi \left(1260 -2625
   \pi ^2+344 \pi
   ^4\right)}{315 \gamma ^7} +\mathcal{O}(\gamma^{-8}),
\end{align}

\begin{align}\label{eprecise}
e(\gamma)=&\frac{\pi^2}{3} -\frac{4 \pi ^2}{3 \gamma } +\frac{4 \pi ^2}{\gamma ^2} -\frac{32 \pi ^2 \left(15-\pi
   ^2\right)}{45 \gamma ^3} -\frac{16 \pi ^2 \left(-15+4 \pi
   ^2\right)}{9 \gamma ^4} +\frac{32\pi^2 \left(-210 +140
   \pi ^2-3 \pi ^4\right)}{105
   \gamma ^5}\notag\\ & -\frac{16 \pi ^2
   \left(-1260+1680 \pi ^2-131
   \pi ^4\right)}{135 \gamma
   ^6}- \frac{256 \pi ^2
   \left(1260-2940 \pi ^2+539
   \pi ^4-6 \pi ^6\right)}{945
   \gamma ^7} \notag\\ &+\frac{256 \pi ^2
   \left(1575-5880 \pi
   ^2+2065 \pi ^4-79 \pi
   ^6\right)}{525 \gamma ^8} +\mathcal{O}(\gamma^{-9}).
\end{align}

The above quantities for $K$ and $m/m^*$ are obtained from the excitation spectrum, which we do not give explicitly here beyond Eq.~(17) of the main text because of a rather lengthy expression. We note that the ground state energy can be connected to, e.g., the sound velocity by the expression from Ref.~\cite{supllieb1963excitations} (that contains a typo corrected here),
\begin{align}
v=\frac{\hbar n}{m} \sqrt{3e-2\gamma \frac{\dif e}{\dif\gamma} +\frac{1}{2}\gamma^2\frac{\dif^2 e}{\dif\gamma^2}}.
\end{align}
Using the exact relation $K=\pi\hbar n/mv$, we obtain another exact relation
\begin{align}
K=\frac{\pi}{\sqrt{3e-2\gamma \frac{\dif e}{\dif\gamma} +\frac{1}{2}\gamma^2\frac{\dif^2 e}{\dif\gamma^2}}}.
\end{align}
Our expressions (\ref{Kprecise}) and (\ref{eprecise}) satisfy the previous relation. We also checked that $Q\lambda=\gamma n$. Finally, using another exact relation for the Lieb-Liniger model given by Eq.~(19) in the main text, we verified the consistency between Eqs.~(\ref{mprecise}) and (\ref{Kprecise}).

We finally note that the ground state energy that is contained in Eq.~(\ref{eprecise}) enables us to find highly accurate expression for the two-particle local correlation function $
g_2={\langle \psi^\dagger\psi^\dagger\psi\psi\rangle}/{n^2}$ at strong interaction. Using the Hellmann-Feynman theorem to reexpress $g_2$ as \cite{gan} $g_2=\dif e/\dif\gamma$, we find
\begin{align}
g_2=&\frac{4\pi^2}{3\gamma^2}\biggl[1-\frac{6}{\gamma }+\frac{8
   \left(15-\pi ^2\right)}{5
   \gamma ^2}+\frac{16
   \left(-15+4 \pi ^2\right)}{3
   \gamma ^3} -\frac{8 \left(-210+140 \pi ^2-3
   \pi ^4\right)}{7 \gamma ^4} +\frac{8 \left(-1260+1680 \pi   ^2-131 \pi ^4\right)}{15
   \gamma ^5}\notag\\
   & +\frac{64 \left(1260-2940 \pi
   ^2+539 \pi ^4-6 \pi
   ^6\right)}{45 \gamma ^6} -\frac{512 \left(1575-5880 \pi
   ^2+2065 \pi ^4-79 \pi
   ^6\right)}{175 \gamma ^7}+\mathcal{O}(\gamma^{-8})\biggr].
\end{align}
It is worth to notice that the first three terms in brackets of the previous equation are found in Ref.~\cite{kormos} using a different technique. Here we revealed several more subleading terms in $g_2$.

\end{document}